\documentclass{article}

\usepackage{arxiv}
\usepackage{graphicx}
\usepackage[T1]{fontenc}
\usepackage[utf8]{inputenc}
\usepackage[english]{babel}
\usepackage{lmodern}
\usepackage{times}
\usepackage{hyperref}       % professional-quality tables
\usepackage{amsmath}
\usepackage{amsthm}
\usepackage{amsfonts}       % blackboard math symbols

\graphicspath{{figures/}}

\newtheorem{definition}{Definition}

\title{Machine Learning-based Positioning using Multivariate Time
  Series Classification for  Factory Environments}
\author{%
	\href{https://orcid.org/0000-0000-0000-0000}{\includegraphics[height=0.8em]{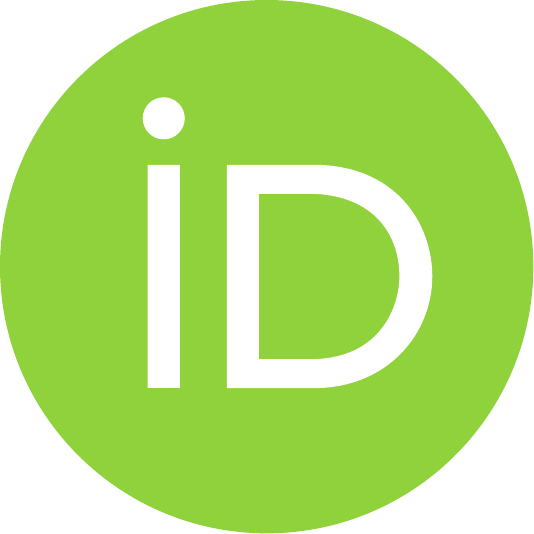}\hspace{1mm}Nisal Hemadasa Manikku Badu} \\
  Institute of Telematics \\
  Hamburg University of Technology \\
  21073 Hamburg, Germany \\
  \texttt{nisal.hemadasa@tuhh.de} \\
	\And
	\href{https://orcid.org/0000-0002-3586-5878}{\includegraphics[height=0.8em]{orcid}\hspace{1mm}Marcus Venzke} \\
  Institute of Telematics \\
  Hamburg University of Technology \\
  21073 Hamburg, Germany \\
	\texttt{venzke@tuhh.de} 
	\And
  \href{https://orcid.org/0000-0001-9964-8816}{%
    \includegraphics[height=0.8em]{orcid}%
    \hspace{1mm}%
    Volker Turau%
  } \\
  Institute of Telematics \\
  Hamburg University of Technology \\
  21073 Hamburg, Germany \\
  \texttt{turau@tuhh.de} \\
	\And
	\href{https://orcid.org/0000-0003-0718-6322}{\includegraphics[height=0.8em]{orcid}\hspace{1mm}Yanqiu Huang} \\
        Faculty of Electrical Engineering\\
	University of Twente\\
	7522NH  Enschede,
The Netherlands \\
	\texttt{yanqiu.huang@utwente.nl} 
      }
      \renewcommand{\shorttitle}{Machine Learning-based Positioning using
        Multivariate Time Series Classification}

\begin{document}

\maketitle

\begin{abstract}
  Indoor Positioning Systems (IPS) gained importance in many
  industrial applications. State-of-the-art solutions heavily rely on
  external infrastructures and are subject to potential privacy
  compromises, external information requirements, and assumptions,
  that make it unfavorable for environments demanding privacy and
  prolonged functionality. In certain environments deploying
  supplementary infrastructures for indoor positioning could be
  infeasible and expensive. Recent developments in machine learning
  (ML) offer solutions to address these limitations relying only on
  the data from onboard sensors of IoT devices. However, it is unclear
  which model fits best considering the resource constraints of IoT
  devices. This paper presents a machine learning-based indoor
  positioning system, using motion and ambient sensors, to localize a
  moving entity in privacy concerned factory environments. The problem
  is formulated as a multivariate time series classification (MTSC)
  and a comparative analysis of different machine learning models is
  conducted in order to address it. We introduce a novel time series
  dataset emulating the assembly lines of a factory. This dataset is
  utilized to assess and compare the selected models in terms of
  accuracy, memory footprint and inference speed. The results
  illustrate that all evaluated models can achieve accuracies above
  80\%. CNN-1D shows the most balanced performance, followed by MLP.
  DT was found to have the lowest memory footprint and inference
  latency, indicating its potential for a deployment in real-world
  scenarios.
\end{abstract}
 \keywords{Indoor positioning \and Machine learning \and Sensor fusion \and Multivariate time series classification}
\section{Introduction}
\label{sec:intro}

Indoor Positioning is a technology widely adopted in many industries,
including medical, sales, manufacturing, logistics and construction
\cite{Farahsari_2022,Li2020}. It is also among the foremost in
technological fronts such as Smart Cities, Industrial Internet of
Things (IIoT) \cite{Frank2022}. In each of these fields, IPS's play
important roles in tracking, navigation, proximity, and inertial
measurements \cite{Farahsari_2022}, thereby injecting more efficiency,
accuracy, and safety to processes.

Our work is motivated by insights from animal behavioral scientists.
Many animal species possess a natural ability to navigate and
recognize their location by utilizing various cues such as geomagnetic
fields, celestial bodies, wind direction, temperature, scent, and
visual landmarks. They develop mental maps through learning and
memory, enabling them to find routes, recognize environments, and
differentiate between different locations. This concept can be applied
to the localization of entities following a predetermined path. By
processing sensory inputs acquired at a given moment or over a
specific time period, an estimation of the current position can be
derived. This could be perceived as a sub-problem of Indoor
Positioning. However, unlike the conventional indoor localization
approaches on determining precise x-y coordinates, we reframe the
problem to ascertain a relative segment on a pre-determined path.

A plethora of research efforts addresses the indoor positioning
problem from a broad range of approaches. These approaches provide
precise x-y coordinates of location estimation. Numerous of these
methodologies heavily depend on external infrastructure for reliable
and robust functionality, such as in RSSI-based solutions that require
consistent signal coverage. However, in the context of applications
involving relative positions rather than precise coordinates, such as
tracking work-in-progress goods in an assembly line, x-y coordinates
bear less significance. Furthermore, deployment of supplementary
infrastructure incur additional cost and can be technically unfeasible
when considering scenarios such as tunnels or mining sites. Secondly,
infrastructure-less systems, like vision-based methods, raise privacy
concerns. Thirdly, some approaches rely on external predetermined
information and cause accumulated errors, such as a starting point in
dead reckoning, which are undesirable in automated processes. Finally,
some approaches operate on non-generic assumptions such as Gaussian
noise or linear motion dynamics. A more detailed comparison between
these existing approaches and our proposed method is contained in
subsequent sections.

In order to overcome the aforementioned limitations, this research
investigates how indoor positions can be learned from sensor data to
enable the execution of the ML model on low power, low performance
devices such as microcontrollers. To this end, we leverage a
combination of inertial sensors (accelerometer, gyroscope, and
magnetometer) and ambient sensors (pressure, temperature, humidity,
and spectrum). The recordings from these sensors form multivariate
time series, which are fused to derive accurate estimates of an
entity's location along a predefined path. The indoor positioning task
is therefore formulated as a Multivariate Time Series Classification
(MTSC) problem. Machine learning (ML) is used to extract the underling
information of the sensor data without making any assumptions about
noise or motion dynamics, and without relying on prior information to
achieve accurate functionality. The challenges arise from the
resource-constraint hardware and maintaining precision, especially
when environmental conditions are changing.

The addressed problem can be more specifically described using an
assembly line in a factory. Assembly lines consist of predetermined
routes planned inside the facility to optimize lead time and cost.
Goods under work-in-progress that traverse along these routes require
tracking in real-time. Modern production lines are digitized and
often use enterprise resource planning (ERP) systems. Especially in
industry 4.0, monitoring of this process is automated. This requires
an asset localization solutions inside highly confidential factory
environments complying with strict privacy policies.

To this end, we evaluate the usage of computationally-light ML models
such as decision trees (DT) \cite{quinlan1986_decision_trees}, RF
\cite{Ho1995_random_forests}, which are well suited for severely
resource constrained edge devices such as low-end Microcontroller
Units (MCUs) with less than 10kB SRAMs \cite{Kumar2017}. We also
choose state-of-the-art of time series classification benchmark
architectures as baselines namely, Multilayer Perceptrons (MLP),
Convolutional Neural Networks (CNN) \cite{IsmailFawaz2019}. Further,
we experiment with the use of more complex Long Short-Term Memory
networks (LSTM) in indoor positioning formulated as time series
classification \cite{Yu2021}.

To the best of our knowledge, this work becomes the first of its kind
in formulating indoor positioning as an MTSC problem by fusing motion
and ambient sensors with no x-y coordinates for particularly relevant
environments such as factories, and investigating applicability of
potential ML models given the limited hardware constraints, to
estimate the relative position. Moreover, state-of-the-art
benchmarking Deep Neural Network (DNN) models for Time Series
Classification (TSC) \cite{Wang_2017,IsmailFawaz2019} are used as
baselines for the models formulated in this work, using a novel indoor
positioning time series dataset. The contributions of this work can be
summarized as follows.

\begin{itemize}
\item Motion and ambient sensor measurements are employed to localize
  a moving entity on a known path consisting of both indoor and
  outdoor components, with no external infrastructure required. The
  importance and drawbacks of the sensors are discussed.
\item A novel multivariate time series dataset is presented. The
  dataset contains sensor measurements from IMU, pressure,
  temperature, humidity and spectrum sensors, collected through
  traversing three paths.
\item The localization problem on a known path is formulated as an
  MTSC problem. ML models, namely DT, RF, MLP, CNN, LSTM are applied
  to solve the MTSC. Their performances are compared against the
  baseline models, based mainly on accuracy, memory footprint,
  inference latency.
\end{itemize}

The rest of this paper is organized as follows. The review on the
related works is in section \ref{sec:related_work}.
Section~\ref{sec:problem_definition} generalizes the indoor
positioning task as an MTSC problem. Section~\ref{sec:ml} introduces
the ML models evaluated in this work. Section~\ref{sec:tuhh_dataset}
describes the novel multivariate time series dataset and a case study
for the proposed solution. Section~\ref{sec:results} presents the
results of evaluations of the formulated ML models and analyses them.
Finally, Section~\ref{sec:conclusion} provides the conclusion of the
work.

\section{Related Work}
\label{sec:related_work}
The existing work on indoor positioning
\cite{Hayward2022,Obeidat2021,Zafari_2019,Yang2021,Wu2018,Ouyang2022,Sesyuk2022,Mostafa2022,Poulose2019,Pascacio2021}
can be categorized along three different dimensions: technology-wise,
technique-wise and algorithm-wise. Technology-wise solutions branch
mainly to satellite-based, radio communication-based, visible
light-based, inertial navigation-based, magnetic-based, sound-based
and vision-based. The majority of solutions within this available
spectrum fail in terms of privacy preservation and independence from
external infrastructure, such as setting up wireless access points
(APs). Although there exist wireless technology based positioning
systems that uphold privacy \cite{holcer2020}, to the best of our
knowledge, all of them require the establishment of supplementary
infrastructure within the localization environment. Moreover, certain
applications, such as tracking goods in assembly lines, necessitate a
relative location of goods rather than precise x-y coordinates.
However, to the best of our knowledge, the primary focus of the
aforementioned solutions is determining x-y coordinates.

Collaborative use of different technologies also exists (Collaborative
Indoor Positioning Systems (CIPS) \cite{Pascacio2021}). Magnetic
field-based localization solutions are independent of the external
infrastructure and also privacy secured. For most of the localization
environments they provide stability, and uniqueness in magnetic
signals \cite{chiang2021}. They also use fusion of several sensors.
However, magnetic based solutions are vulnerable to dynamic
environments where electromagnetic disturbances occur such as the
motion of metallic structures close to the magnetometer.

Work classified under technique-wise includes dead-reckoning-based,
vision analysis-based, triangulation-based, fingerprinting-based and
proximity-based. To the best of our knowledge, the latter three
techniques require external infrastructure for their operation.
Dead-reckoning, to the best of our knowledge, require external
information such as initial location and is prone to accumulated
error. Vision based systems pose a significant risk to privacy.

Algorithm-wise classification mainly branches to the least square
method, maximum likelihood method, deterministic or probabilistic
method \cite{Pascacio2021}. Existing fusion-based positioning
methods range from conventional methods such as least squares, maximum
likelihood, maximum a posterior, and minimum mean squares error, to
state estimate methods such as hidden Markov model, Kalman filter,
extended Kalman filter, and particle filter, and ML methods such as
k-nearest neighbors, random forests (RF), support vector machine, and
neural networks \cite{Guo2020}. The use cases of ML in this regard
have so far been mainly applied to positioning systems based on
Received Signal Strength (RSS) and fingerprint \cite{Guo2020,
  Nessa2020}. Furthermore, they mainly assume Gaussian noise and
linear motion dynamics \cite{Nessa2020}, which may not describe
real-world positioning systems.

\section{Problem Definition}\label{sec:problem_definition}

Indoor positioning is heavily used in asset localization and
work-in-progress tracking in production lines, factories and
warehousing. The goal is to accurately determine an asset's location
within an enclosed facility as the assets lie stationary or move along
predetermined paths. By dividing these paths into smaller segments,
the location of the asset, at a given time, can be determined by
finding the most likely segment the asset is present in. This can be
interpreted as a classification problem. As the asset moves along the
path, motion and ambient sensors attached to the asset acquire
measurements and record them periodically, in resulting a multivariate
time series. This multivariate time series data is then used as inputs
to the classification problem on the edge device, on the fly. This
problem can be defined in general terms as follows.

\begin{definition}
  Let $P$ be a path partitioned into $l$ segments, such that
  $P\rightarrow [s_{1},\ldots, s_{m},\ldots,s_{l}]^{T}$. For each
  $m\in \{1,\ldots,l\}$ segment $s_{m}$ is uniquely identified by a
  label $y_{m}$ from a set
  $Y_{P} = \{y_{1}, y_{2}, ... , y_{m}, ... , y_{l}\}$ of labels of
  $P$.
\end{definition}
%{\medskip}
An asset is being carried along the path $P$. Neither the speed nor
the variation of the speed along the path are known. Further, the
times of completion of traversal along the path do not remain constant
across several instances of path traversals. Hence, no trivial
correlation between asset's position and the time exists. Further, the
path $P$ contains segments along which the asset is carried both on
its forward and return journeys, which is analogous to the asset
turning around and returning on the same route it took in the forward
journey. However, it does not include anomalies such as long pauses
while traversing or moving out of the defined path, which could be
typically expected in industrial environments such as factories or
warehouses.

The following two definitions formally capture the notions
of {\em observation} and {\em multivariate time series} for paths.

\begin{definition}
  Let $X^{i}$ denote a univariate time series of a feature $i$,
  engineered from the recordings of sensor measurements, as a result
  of traversing a complete path $P$. An observation at a given
  sampling time $t$ is denoted as $x_{t}^{i}$ and $X^{i}=[x_{1}^{i},
  x_{2}^{i},...,x_{k}^{i}]^{T}$, where $k$ is the total number of
  observations of time series $X^{i}$.
\end{definition}

\begin{definition} 
  Let there be $n$ different features, giving distinct univariate time
  series for a data-collection run along $P$. Then,
  $X=[X^{1}, X^{2},...,X^{n}]$ is then defined as a multivariate time
  series for $P$.
\end{definition}

Finally we formally define the problem addressed in this work.

\begin{definition}
  Let $X_{t} = [x_{t}^{1}, x_{t}^{2},..., x_{t}^{n}]$ be the
  observations of all features in the feature space, at a given time
  $t$. The problem addressed in this work is to find a function
  $f_{P}^{j}:\{X_{t}, X_{t-1},...,X_{t-j}\} \rightarrow Y_{P}$, that
  determines the label $y_{m}$ of the segment in which the object is
  at time $t$, using time series $[X_{t}, X_{t-1},...,X_{t-j}]$, where
  $j$ is a predetermined window size, such that
  $j\in \mathbb{Z}_{0}^{+}$, and for $j \leq t - 1$ it holds that
  $f_{P}^{j}(X_{t}, X_{t-1},...,X_{t-j}) = y_{m}$.
\end{definition}

\section{Indoor Positioning using DT, RF, CNN and LSTM Networks}
\label{sec:ml}
In this section, we formulate the architectures of the ML models that
we use to learn the function $f_{P}^{j}$ defined in definition 4. We
use two recently popular time series classification baseline models
namely, MLP, and Fully Convolutional Networks (FCN)
\cite{IsmailFawaz2019, Wang_2017}. Additionally, we use a tree-based
approach, namely DT with entropy \cite{Quinlan1986,Salzberg1994},
which is relatively less complex and therefore suitable for use on
low-performance edge devices, such as MCUs. We also explore an
ensemble approach, namely RF \cite{Ho1995_random_forests,Breiman2001}.
Lastly, we apply the dataset to vanilla LSTM \cite{Hochreiter1997},
bidirectional LSTM (BiLSTM), CNN-1D and CNN-2D
\cite{lecun1995convolutional} for solving the time series
classification task.

The runtime environment for most existing works related to indoor
positioning is a resource-constrained edge device \cite{Nessa2020}.
Therefore, when optimizing ML models, it is not only important to
consider the accuracies but also their resource usage. Hence, in this
work, we evaluate less complex and lighter models with significant
accuracy to fit the problem defined in section
\ref{sec:problem_definition}.

\subsection{Decision Trees: DT}
\label{sub:decision_trees}
In this work, the parameters used for DT are the split criterion (entropy/gini) and the maximum depth (8-22). The best accuracy performance on average is obtained with entropy with a maximum depth of 14, across all three paths. The influence of other hyperparameters such as the minimum sample split and minimum samples leaf, on the accuracy score, is found to be insignificant. Hence maximum tree depth is the only early stopping criteria used. Furthermore, adding more features beyond the most important 17 features does not noticeably improve the model. Consequently, the 17 features mentioned in subsection \ref{sub:data_preprocessing} are chosen as input features for all model architectures used in this work. 

\subsection{Multilayer Perceptrons}
\begin{figure}[htbp]
	\centering
	\includegraphics[scale=0.8]{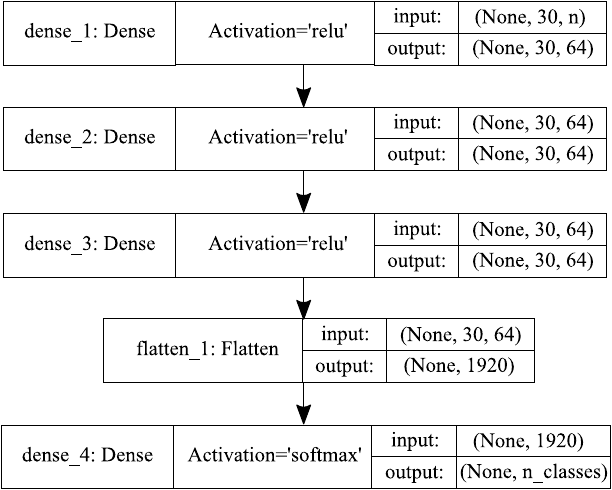}
	\caption{The model architecture of the MLP. $n$ and $n\_classes$ represent the number of input features and the number of output classes respectively.}
	\label{fig:mlp}
\end{figure}

A basic MLP is used as a baseline model as proposed by Wang et al.
\cite{Wang_2017}. We fine-tune the model with minor modifications to
the hypterparameters to fit our dataset (see Fig~\ref{fig:mlp}). An
input layer takes 2D inputs with dimensions of (30, 17), (30, 17), and
(30, 9) for paths 1, 2, and 3, respectively. Here, 30 denotes the
window of timesteps of the time series data taken as inputs, and 17,
17, and 9 are the number of input features of the datasets for paths
1, 2 and 3 respectively. Three hidden fully connected layers with 64
nodes and Rectified Linear Unit (ReLU) \cite{nair2010rectified}
activation functions follow. A Flatten layer is included before the
output layer with a softmax activation function. Dropouts
\cite{srivastava2014dropout} are omitted as they result in a decrease
in training, validation, and test accuracies. Categorical cross
entropy is used as the loss function, and the Adam optimization
algorithm \cite{adam_kingma_2015} is used with a learning rate of
0.0003. The batch size is set to 256, and the model is trained over 30
epochs.

\subsection{FCN}
\begin{figure}[htbp]
	\centering
    \includegraphics[scale=0.8]{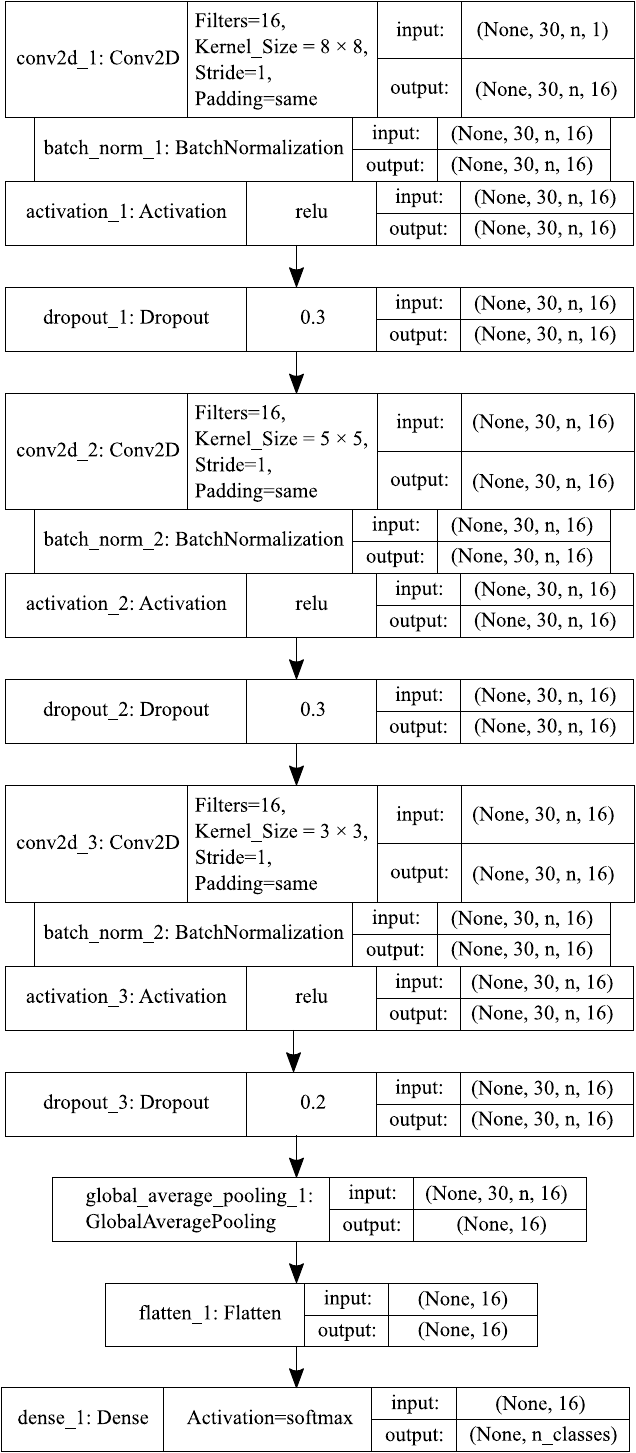}
	\caption{The model architecture of the FCN. $n$ and $n\_classes$ represent the number of input features and the number of output classes respectively.}
	\label{fig:fcn}
\end{figure}

The proposed FCN architecture by Wang et al. \cite{Wang_2017} is used as a baseline. Fig.~\ref{fig:fcn} shows our refined FCN architecture. This model has an input layer taking inputs with dimensions of (30, 17, 1), (30, 17, 1), and (30, 9, 1), for paths 1, 2 and 3 respectively. 
Here, following the same pattern as in the case of MLP, 30 denotes the input sequence length and 17, 17, and 9 are the size of the feature space for paths 1, 2 and 3 respectively. 
The hidden layers are three distinct convolutional blocks, each containing a 2D convolutional layer (Conv2D), batch normalization layer (BN) \cite{Ioffe_2015}, ReLu activation layer. The three convolution layers have 16 filters, 2d-kernels of sizes 8, 5 and 3 and stride 1. To avoid overfitting, dropout values of 0.3, 0.3 and 0.2 are added respectively. The convolutional blocks are followed by a 2D global average pooling layer and a Flatten layer. The output dense layer is activated using a softmax activation function, and the loss function used is the categorical cross entropy with Adam optimization and a learning rate of 0.00005. The batch size is 100 and the model is trained for 30 epochs.

\subsection{Random Forests: RF}
The accuracy performance of RF is evaluated across the decision-tree-estimator range from 15 to 35, in multiples of 5. The maximum depth of an estimator is varied within the range from 12 to 16, in the multiples of 2. When the parameters are set below these ranges, the accuracy performance significantly decreases. Above this range, there is no significant improvement in accuracy, however the memory footprint of the models increases drastically. Entropy is maintained as the splitting criterion of all the cases. Considering these factors, maximum depth of 14 is chosen as the optimum, which is equivalent to the best performance achieved using DT, with 25 decision-tree-estimators

\subsection{Convolutional Neural Networks: CNN}
Using CNN, we optimize two architectures, one using a 1D convolutional
layer and the other using a 2D convolutional layer, followed by dense
layers in both the cases. 

%In the case of CNN-1D, similar to MLP, the inputs of dimensions are set to (30, 17), (30, 17), and (30, 9) for routes 1, 2, and 3, respectively (see Fig.\ref{fig:2d-cnn}). Here, 30 denotes the window of timesteps of the time series data taken as inputs, and 17, 17, and 9 are the number of input features of the datasets for routes 1, 2 and 3 respectively. The first layer is a 1D-convolutional layer (Conv1D) with 64 filters and kernel size as 5, activated by a ReLu function. Its output is then passed to a 1D-max pooling layer (MaxPooling1D) with pool size 3 $\times$ 3, stride size 1. The padding is set to 'same', which means that the output feature map size will be the same as the input feature map size. The output of the max pooling layer is batch normalized and subsequently passed to a Flatten layer. Then follows two fully connected layers each of size 32, both ReLu-activated. The last is a dense layer, connected fully to the previous dense layers, with a softmax activation function. 

\begin{figure}[htbp]
	\centering
	\includegraphics[scale=0.8]{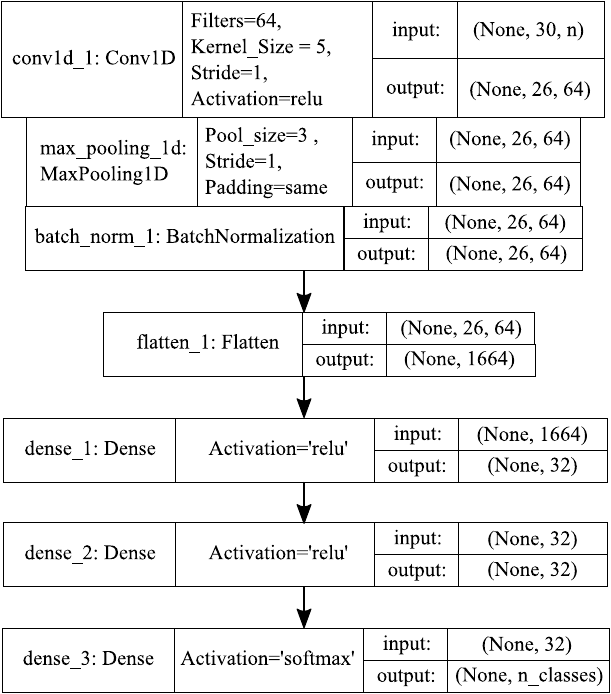}
	\caption{The model architecture of the CNN-1D. $n$ and $n\_classes$ represent the number of input features and the number of output classes respectively.}
	\label{fig:2d-cnn}
\end{figure}

The first CNN model (CNN-1D), similar to MLP, the inputs of dimensions are set to (30, 17), (30, 17), and (30, 9) for paths 1, 2, and 3, respectively. Here, 30 denotes the window of timesteps of the time series data taken as inputs, and 17, 17, and 9 are the number of input features of the datasets for paths 1, 2 and 3 respectively (see Fig.~\ref{fig:2d-cnn}).
The first layer is a 1D-convolutional layer (Conv1D) with 64 filters and kernel
size as 5, activated by a ReLu function. Its output is then passed to
a 1D-max pooling layer with pool size 3$\times$3, stride size 1, with
padding set to `same'. The output is then batch normalized and
subsequently passed to a Flatten layer. Then follows two fully
connected layers each of size 32, both ReLu-activated. The last is a
dense layer, connected fully to the previous dense layers, with a
softmax activation function.

\begin{figure}[htbp]
	\centering
	\includegraphics[scale=0.8]{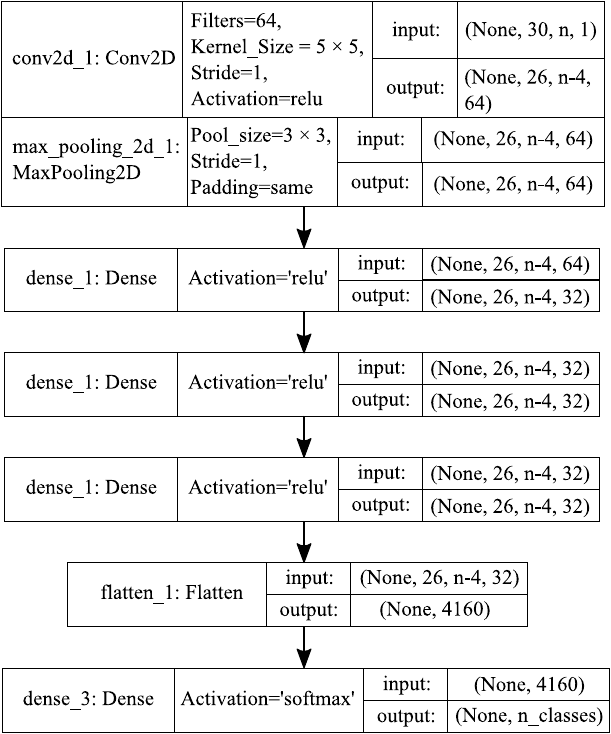}
	\caption{The model architecture of the CNN-2D. $n$ and $n\_classes$ represent the number of input features and the number of output classes respectively.}
	\label{fig:1d-cnn}
\end{figure}

In the second model (CNN-2D), input dimensions are (30, 17, 1), (30, 17, 1), and (30, 9, 1) for paths 1, 2, and 3, respectively, where the 3rd dimension with the value 1 refers to the number of channels (see Fig.~\ref{fig:1d-cnn}). The first layer consists of a
2D-convolutional layer with 64 filters and kernel size 5$\times$5,
with no padding and is activated by a ReLu. A 2D-max pooling layer is
then applied, with pool size 3$\times$3, stride 1 and padding 'same'
type. This is followed by a layer of batch normalization, two fully
connected layers both having ReLU activated, a Flatten layer, finally
leading to a dense layer with softmax activated.

The categorical cross entropy loss function and Adam optimization with a learning rate of 0.00003 is adopted for both cases above. The model is trained for 30 epochs.

 \subsection{Long Short-Term Memory networks: LSTM}
\begin{figure}[htbp]
   \centering \includegraphics[scale=0.8]{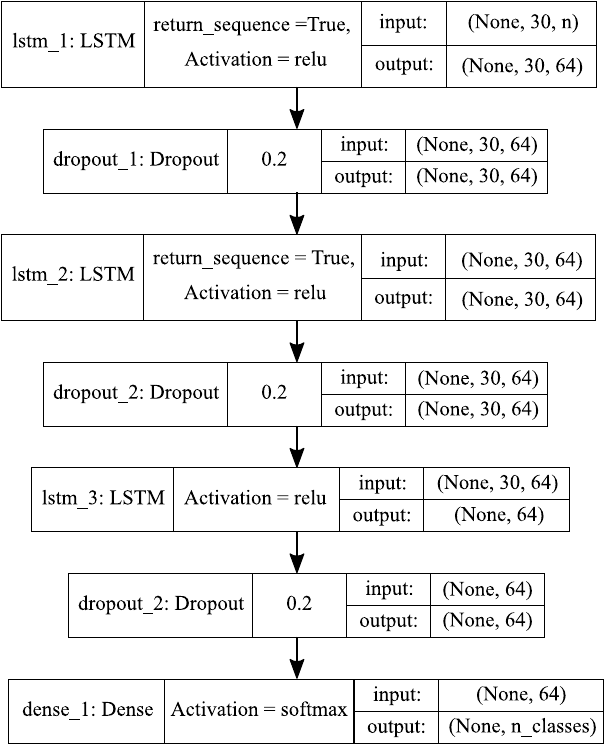}
 	\caption{The model architecture of the LSTM. $n$ and $n\_classes$ represent the number of input features and the number of output classes respectively.}
 	\label{fig:lstm}
 \end{figure}

 The Long-Short Term Memory (LSTM) architecture is implemented using an input layer of dimensions (30, 17), (30, 17), and (30, 9) for paths 1, 2, and 3, respectively (see Fig.~\ref{fig:lstm}). 
Three LSTM layers of size 64 each, are connected via 3 dropout layers,
each of 0.2, sequentially connecting to a dense layer with a softmax
activation function. 
The categorical cross entropy is selected as the loss function and Adam optimization is employed with a learning rate of 0.00003. The model is trained using batch sizes of 200 for 30 epochs.

\subsection{Bidirectional Long Short-Term Memory networks: BiLSTM}
\begin{figure}[htbp]
	\centering
	\includegraphics[scale=0.8]{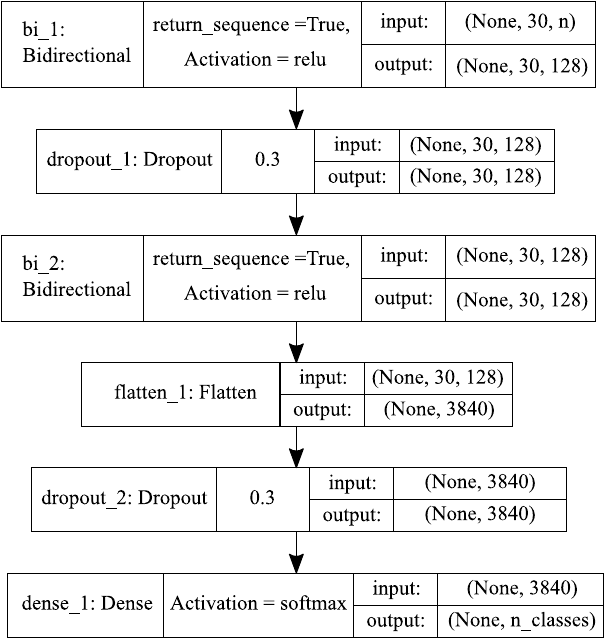}
	\caption{The model architecture of the BiLSTM. $n$ and $n\_classes$ represent the number of input features and the number of output classes respectively.}
	\label{fig:bidirectional}
\end{figure}

The proposed bidirectional-LSTM architecture takes inputs with dimensions (30, 17), (30, 17), and (30, 9) for paths 1, 2, and 3, respectively (see Fig.~\ref{fig:bidirectional}).
Two BiLSTM layers, each of size 64, are connected via dropouts of 0.3,
which leads to the output layer which is activated with a softmax
activation function.
The categorical cross entropy is used as the loss function and Adam optimization is employed with a learning rate of 0.00003. The model is trained with a batch size of 1024 for 30 epochs.

In the models, the dimensions of the input depend on the time step
window size $j$, and on the feature space size $n$ as described in
definition 4 and 3 respectively.

\section{Localizing a Moving Object: Motion-Ambient Dataset, Data
  Preprocessing and Feature Selection}
\label{sec:tuhh_dataset}

In this section, we describe a use case scenario to verify the claims
of this paper. This is done using the \textit{Motion-Ambient} dataset,
which is presented for the first time.

We construct a practical scenario, comprising events, that one can
expect to have in factories and warehousing, inside the premises of
XXX. A moveable data-logging setup is used to collect IMU, pressure,
humidity, temperature, and spectrum data across three different paths
comprising both indoor and outdoor segments. These paths comprise a
variety of dynamics including passages in buildings, elevators, ramps,
stairs, terrain with varying roughness such as cobblestone, areas with
different lighting conditions, magnetic disturbances near metallic
structures (e.g. metal door frames), etc., which are commonly
encountered in industrial settings. These paths are annotated in
real-time with pre-determined segments (or classes) as the setup is
being transported.

Motion-Ambient is a time series dataset that is created to benchmark
research in indoor localization. In the following subsections, we
describe this dataset and its preprocessing steps. Motion-Ambient
dataset will be made public including a detailed description,
visualizations, analysis and benchmarking of it.

\subsection{Dataset Description}\label{sub:dataset_description}

The dataset is mainly focused on indoor localization use cases.
However, the data-collection-paths pass though indoors spaces,
outdoors areas, over diverse terrains, under varying lighting, weather
and climatic conditions. The summary of the statistics are given in
table \ref{table:dataset_summary_1}.

The data-logging-hardware includes two dedicated modules especially
designed for this work, based on Raspberry Pi development boards
integrated with sensors. These sensor types and their constituent
sensor modules are summarized in table \ref{table:sensors}. Outdoor
localization technologies such as GPS and GNSS are not featured among
the hardware. Hence they are not features in the dataset. The sampling
rate of sensors from the IMU (acceleration, rotation and magnetic flux
density) is 24 Hz. The rest of the sensor measurements are sampled at
0.14 Hz. The annotation of the data from Rapsberry Pi's are done by
connecting a smart phone to it via Bluetooth.

\begin{table*}[htbp]
  \caption{Summary of the dataset: data-collection-runs and the
    samples for each path traversed and the sensors involved}
  \begin{center}
    \begin{tabular}{|c|c|c|c|c|}
      \hline
      \textbf{Path}&\textbf{Distance (m)}&\textbf{Runs}&\textbf{Samples}&\textbf{Sensors}\\
      \hline
       1&470&115&1,597,657&IMU measurements, temperature, humidity, pressure, spectrum \\
      \hline
       2&233&180&1,439,903&IMU measurements, temperature, humidity, pressure, spectrum\\
      \hline
       3&327&115&1,597,657&IMU measurements\\
      \hline
    \end{tabular}
    \label{table:dataset_summary_1}
  \end{center}
\end{table*}

\begin{table*}[htbp]
  \caption{Dataset before preprocessing: $t_{i}$ and $t_{j}$ are
    arbitrary sampling times, where $t_{i} \neq t_{j}$. $T$ is the
    sampling period of IMU sensor. `\checkmark' denotes recorded
    sensor measurements. `-' indicates that no measurement acquired.}
  \begin{center}
    \begin{tabular}{|c|c|c|c|c|c|c|}
      \hline
      \textbf{Timestamp}&\textbf{IMU-accelerometer-x}&\textbf{...}&\textbf{IMU-magnetometer-z}&\textbf{temperature}&\textbf{humidity}&\textbf{...}\\
      \hline
      $t_{i}$&\checkmark&...&\checkmark&-&-&...\\
      \hline
      $t_{i+T}$&\checkmark&...&\checkmark&-&-&...\\
      \hline
      $t_{j}$&-&-&...&\checkmark&\checkmark&...\\
      \hline
      $t_{i+2T}$&\checkmark&...&\checkmark&-&-&...\\
      \hline
    \end{tabular}
    \label{table:dataset_before_preprocessing}
  \end{center}
\end{table*}

\begin{table*}[htbp]
  \caption{Sensors used in the data-logging-hardware and the features
    extracted from them.}
  \begin{center}
    \begin{tabular}{|c|c|c|}
      \hline
      \textbf{Sensor module}&\textbf{Sensor types}&\textbf{Features extracted}\\
      \hline
      ICM-20948 from TDK&accelerometer&acceleration in x, y, z components [G]\\
      \cline{2-3}
                            & gyroscope &rotation about x, y, z axes [degrees per second]\\
      \cline{2-3}
                            & magnetometer & magnetic flux density in x, y, z components [$\mu$T]\\
      \hline
      BME 688 from Bosch&temperature sensor&temperature[\textdegree C]\\
      \cline{2-3}
                            &humidity sensor&humidity [\% r.H]\\
      \cline{2-3}
                            &pressure sensor&pressure [Pa]\\
      \hline
      AS7341 from AMS&spectrum sensor&yellow, green, blue, indigo, violet [Counts]\\
      \hline
    \end{tabular}
    \label{table:sensors}
  \end{center}
\end{table*}

\subsection{Dataset Preprocessing}
\label{sub:data_preprocessing}
Table \ref{table:dataset_before_preprocessing} shows the dataset
before the preprocessing stage. The sensors integrated to the
Raspberry Pi have varying and unsynchronized sampling rates. Hence,
records are periodically missing in some data columns, especially from
the ambient sensors with low sampling rate compared to IMU sensors. To
deal with these missing values and to match the sampling rates, in our
approach we firstly perform backward filling (padding the missing
fields with the lastly recorded value) followed by forward filling
(padding the missing fields with the next earliest value that is
recorded to fill the remaining missing values after backward filling).
The resulting data is then applied with a rolling mean filter. The
size of the filter is chosen to be marginally greater than the maximum
consecutive missing values recorded in that particular column. This
operation is performed separately for each column where values are
missing. The data is finally min-max normalized, separately for each
column. Columns from the resulting dataset-table are selected as
features.

\subsection{Feature Selection}
\label{sub:feature_selection}
Out of 7 individual sensors types from 3 sensor modules, 17
measurements are chosen as features. These are given in table
\ref{table:sensors}. The selection is based on the highest feature
importance figures in the DT model. This set of features are thereon
used for paths 1 and 2 consistently throughout the evaluations using
other models described in this work. For path 3, only the 9 IMU sensor
measurements are used as features.

\section{Results, Analysis and Discussion}
\label{sec:results}
In this section, we describe the constraints used in structuring the
architectures of the formulated ML models. Furthermore, we introduce
the metrics used to evaluate the performance of the models. Based on
these metric-scores, we compare, discuss and analyse further insights
in detail.

\subsection{Description of model architectures and their constraints}
\label{sub:architecture_description}
Apart from its use in regression problems, the purpose of decision
trees is to classify a dataset based on its feature characteristics,
while dense layer networks such as MLP can be used for the same tasks,
as well as for identifying complex, non-linear patterns in data. RF
also used for similar use cases, generally with a higher accuracy and
unbiased predictions than decision trees, at the cost of more
computations and memory utility. Originally, Convolutional Neural
Networks (CNNs) are used to identify spatial patterns in images.
However, multivariate time series data can be arranged as a 2D heat
map similar to an image. This enables the use of CNNs to capture the
local temporal patterns, across time and features. By combining CNNs
with dense layers, the aim is to reduce the image patterns learnt into
a simple classification problem. Furthermore, CNNs generate large
feature spaces. Complex patterns and inter-correlations among these
large feature sets can be learnt using the dense layers. Long
Short-Term Memory (LSTM) models, on the other hand, learn long-term
dependencies between time steps in time series or sequence data. That
is, while CNNs are capable of recognizing patterns of a local region
of focus of a multivariate time series, LSTMs are capable of learning
the relationships between and among several of these regions. In
contrast DTs, RFs and MLPs, compared to LSTMs are not specially
designed to extract time correlations of data. However, feeding a time
window of data has a positive effect on the accuracy, implying that
they are capable of detecting patterns across time, up to an extent.

The presented DT model is optimized after an architecture search with
variables namely, split criterion (entropy/gini) and the maximum depth
(8-22). The best accuracy performance on average is obtained with
entropy, with a maximum depth of 14, across all three paths. The
influence of other hyperparameters such as the minimum sample split
and minimum samples leaf, on the accuracy score, is found to be
insignificant. Hence maximum tree depth is the only early stopping
criteria used.

Before selecting the presented architecture of RF, accuracy
performance of RF is evaluated across the estimators ranging from 15
to 35, in multiples of 5. The maximum depth of an estimator is varied
within the range from 12 to 16, in the multiples of 2. The optimum
results are obtained with a maximum depth of 14, with 25
decision-tree-estimators.

We consider MLP and FCN models proposed by Wang et al. (2017)
\cite{Wang_2017} to benchmark the neural network models that we
present. Having them as a reference, we experiment with how CNN and
LSTM model architectures and their variations to solve the indoor
positioning problem. The variation of CNN architecture includes
CNN-1D, and CNN-2D, whereas in LSTM, vanilla LSTM (LSTM) and BiLSTM.
Hence the architectures of the latter mentioned models are not allowed
to vastly deviate from the benchmark models. MLP and FCN models use
three blocks of layers in their architecture, except for softmax
activated final dense layer. Hence in the variants CNN-1D and LSTM we
use architectures with three blocks of layers. An additional dense
layer is added of CNN-2D and one BiLSTM layer is taken out from BiLSTM
to improve the performance in terms of comparison metrics to an
acceptable level compared to benchmark models.

The timestep window size $j=30$ yields the highest accuracy out of
\{10, 20, 30, 40, 50\}, for the benchmarking models MLP and FCN. Above
this value, the models tend to overfit, and below this value, the
models underfit. We use the same timestep window size for all models.

\subsection{Accuracy Metrics}
In this work, in addition to the accuracy-score used in ML, we use
another accuracy metric fine tuned to this specific application called
Loc-score.

\subsubsection{Accuracy-score}
ML classification accuracy-score (accuracy) is the ratio of the number
of correct predictions to the total number of predictions, as shown by
equation \ref{eq:accuracy}. In this work, this is equivalent to the
proportion of sensor samples correctly classified with no disparity
against annotations that they are assigned.

\begin{equation}
  \textrm{Accuracy-score}=\dfrac{\textrm{Total number of correct predictions}}{\textrm{Total number of predictions}}
  \label{eq:accuracy}
\end{equation}

However, there is no guarantee that the labels used in this work are
in 100\% agreement with the ground truth, due to labelling noise. This
is mainly attributed to the high sampling rates of the sensors which
are much larger than average human reaction times, making it difficult
to consistently annotate the transitions between segments of a path
for all data-collection-runs. To compensate for such inconsistencies
in labelling, Grewe introduces the accuracy metric loc-score
\cite{Grewe2021}.

\begin{table*}[htbp]
  \caption{Accuracy-scores across the ML models. The highest accuracy
    corresponding to each path is highlighted. }
  \begin{center}
    \begin{tabular}{|c|c|c|c|c|c|c|c|c|}
      \hline
      \textbf{Path}&\textbf{MLP}&\textbf{FCN}&\textbf{Decision Tree}&\textbf{RF}&\textbf{ LSTM}&\textbf{BiLSTM}&\textbf{CNN-1D}&\textbf{CNN-2D}\\
      \hline
       1&0.8848&0.8301&0.8394&0.8977&0.8135&0.8445&\textbf{0.9105}&0.8635\\
      \hline
       2&0.9520&0.9298&0.8405&0.8946&0.8735&0.8872&\textbf{0.9544}&0.8891\\
      \hline
       3&\textbf{0.9321}&0.9185&0.8777&0.9217&0.9116&0.9013&0.9302&0.8939\\
      \hline
    \end{tabular}
    \label{table:accuracy}
  \end{center}
\end{table*}

\begin{table*}[htbp]
  \caption{Loc-scores across the ML models. The highest loc-score
    corresponding to each path is highlighted. }
  \begin{center}
    \begin{tabular}{|c|c|c|c|c|c|c|c|c|}
      \hline
      \textbf{Path}&\textbf{MLP}&\textbf{FCN}&\textbf{DT}&\textbf{RF}&\textbf{LSTM}&\textbf{BiLSTM}&\textbf{CNN-1D}&\textbf{CNN-2D}\\
      \hline
       1&0.9079&0.8542&0.8529&0.9199&0.8270&0.8205&\textbf{0.9315}&0.8367\\
      \hline
       2&0.9660&0.9465&0.854&0.9108&0.8932&0.9058&\textbf{0.9679}&0.9058\\
      \hline
       3&\textbf{0.9438}&0.9312&0.9033&0.9350&0.9244&0.9164&0.9416&0.9057\\
      \hline
    \end{tabular}
    \label{table:locscore}
  \end{center}
\end{table*}

\subsubsection{Loc-score}
Loc-score defines a window of timesteps around transitions from one
segment (considered as a class in this problem) to the next, in the
true class labels. During the evaluation, predictions to either of the
two classes, within this window, are considered correct, while
prediction to other classes are considered misclassified. The ratio of
the samples consequently correctly predicted and the total number of
predictions is defined as the Loc-score. This can be more formally
defined as follows.

\begin{definition}
  For a transition from a segment $y_m$ to $y_{m+1}$, at a given
  timestep $t_{tr}$, and a defined window size $2\tau + t_{tr}$, a
  classification $\hat{y}_{t}$, at time $t$, is considered correct
  only if $\hat{y}_{t} \in \{y_{m}, y_{m+1}\}$, such that
  $t\in [t_{tr}-\tau, t_{tr}+\tau]$. Then,
	\begin{equation*}
          \textrm{Loc-score}=\dfrac{\textrm{number of correct predictions (per definition)}}{\textrm{Total number of predictions}}
	\end{equation*}
\end{definition}

\subsubsection{Memory Footprint}
The memory footprint of a trained ML model is the amount of memory required to store the network's parameters, including the network structure, the trained weights and biases of all layers. The higher the memory footprint, the more resources are required by the hardware to deploy the network.

\subsubsection{Inference Latency}
Inference latency is the time taken for a ML model to make a prediction or classification based on input data. 

\subsubsection{Throughput}
Throughput is defined as the rate of predictions, which in our case is the number of predictions per ms. 

Both the inference latency and throughput are measures of how quickly
a model can process new data and make accurate predictions.

\subsection{Analysis of Results}
The results are analysed and discussed in this subsection.

\subsubsection{Accuracy}
Table~\ref{table:accuracy} and Table~\ref{table:locscore} demonstrate
that the CNN-1D model exhibits the highest accuracy and loc-score
numbers for both path 1 and 2. For path 1, CNN-1D is closely followed
by the RF and MLP models, respectively. MLP model performs best in
terms of accuracy for path 3, followed by the CNN-1D model marginally,
and RF with a distinct separation. For path 2, the accuracy value of
RF is significantly lower than those of MLP and CNN-1D, and even lower
than that of FCN. FCN, BiLSTM, LSTM, and CNN-2D follow thereafter,
with their rankings fluctuating for each path. In average, DT has the
lowest accuracy, with the exception of path 1.

For $j=1$, DT and RF both yielded significantly lower accuracy-scores
(0.7737, 0.7873, 0.8264 and 0.8216, 0.8722, 0.8583 respectively) and
loc-scores (0.7959, 0.7981, 0.8422 and 0.8483, 0.8901, 0.8732
respectively) for paths 1, 2 and 3, compared to when $j=30$. This
demonstrates that tree-based models can capture time correlations to a
certain degree, even though they are not tailored for this. Accuracy
of DT and RF could be further increased by engineering more optimal
features.

In summary, architectures such as MLP, CNN-1D and RF that extract
short-to-mid range time dependencies yield higher accuracy figures.
The fact that two LSTM variants do not lead to a significant boost in
classification accuracy could mean that the dataset does not contain a
higher amount of important long-term patterns these models can
identify. For paths 2 and 3, the temporal correlation extraction
capability of DT is not enough to capture the same amount of feature
dynamics as the other models. However, the dataset for path 1 has less
complex, more structured data making it easier for both DT and RF to
capture. This can be concluded by the relative increment in accuracy
metric values for DT and RF in path 1 compared to the other two paths.

CNN-1D demonstrates higher accuracy values than CNN-2D for all three
paths. This could be due to architectural changes, such as the
addition of batch normalization between layers, which helps regularize
the model and improve accuracy in CNN-1D. Additionally, it is possible
that Conv1D extracts more features that better represent the data than
Conv2D. Conv1D convolves along feature vectors, arranged according to
the temporal sequence, which could be more effective at deriving
informative features than convolving only a part of the feature space
along with their temporal dynamics, all at once.
 
Despite LSTM and BiLSTM being architectures capable of extracting
complex temporal correlations, they do not produce the best results.
To understand why, initially, the hyperparameters are varied within
the limits of constraints mentioned in subsection
\ref{sub:architecture_description}, that is, a maximum of 3 layers
with 64 cells in each layer and $j=30$. This does not improve
accuracy. In conclusion, we are left with several reasons such as the
dataset having simple temporal complexities, the amount of data
required for the models to learn being simply too small, the time step
window size $j=30$ being not large enough, the lesser number of layers
results in a too shallow network in the case of BiLSTM, etc., which we
do not cover in this paper.

\begin{table*}[ht]
  \caption{Memory footprint of the ML models in MB. The best memory
    footprint corresponding to each path is highlighted.}
  \begin{center}
    \begin{tabular}{|c|c|c|c|c|c|c|c|c|}
      \hline
      \textbf{Path}&\textbf{MLP}&\textbf{FCN}&\textbf{DT}&\textbf{RF}&\textbf{LSTM}&\textbf{BiLSTM}&\textbf{CNN-1D}&\textbf{CNN-2D}\\
      \hline
       1&7.72&18.72&\textbf{0.99}&24.65&30.92&75.96&5.37&67.00\\
      \hline
       2&7.63&18.72&\textbf{0.44}&11&30.87&75.74&5.36&66.50\\
      \hline
       3&7.65&9.93&\textbf{0.72}&17.98&30.88&75.78&5.35&25.65\\
      \hline
    \end{tabular}
    \label{table:memory_footprint}
  \end{center}
\end{table*}

\subsubsection{Memory Footprint}
Table \ref{table:memory_footprint} shows that, the lowest memory
footprint is consumed by the DT, for all three paths. These values are
significantly lower than the rest. The highest memory requirement is
demanded by the BiLSTM model. For path 3, the memory footprints of the
FCN and CNN-2D models are lower than the their corresponding versions
of paths 1 and 2. This is mainly attributed to the smaller size of the
input feature space for path 3 (9), as compared to the other paths
(17), since FCN and CNN-2D models depend on the size of the input
feature space.

The memory footprint of DT models can be readily accommodated by
high-end MCUs such as the ESP32 \cite{ESP32}, which has an average of
500kB SRAM and 4MB of flash memory, without further compression, using
technologies such as swapping \cite{swapping2021}. Pruning of these
architectures can reduce their size further, enabling them to be
deployed in severely resource-constrained MCUs with minimal loss of
accuracy \cite{Kumar2017}. In edge ML, the device's size impacts more
dominantly than the model architecture, on the energy budget
\cite{micronets2021}. This can reduce energy consumption and make them
suitable for long-term applications, matching well expectations from
the industrial positioning applications. Other DNN models can also be
deployed on high-end MCUs after optimizing them to fit the
resource-constrained hardware
\cite{banbury2021mlperf,fedorov2020tinylstms}. However, this could
lead to accuracy compromises.

%\begin{figure}[htbp]
%	\centering
%	\includegraphics[scale=0.8]{plot_memory_footprint_routes.pdf}
%	\caption{Memory footprint of the machine learning models in MB.}
%	\label{fig:memoryfootprint}
%\end{figure}
%
%
%\begin{figure}[htbp]
%	\centering
%	\includegraphics[scale=0.8]{plot_infer_time_routes.pdf}
%	\caption{Inference latency of different machine learning models with standard deviation in ms.}
%	\label{fig:infer_time}
%\end{figure}

\begin{figure*}[ht]
  \centering
  \includegraphics[scale=0.95]{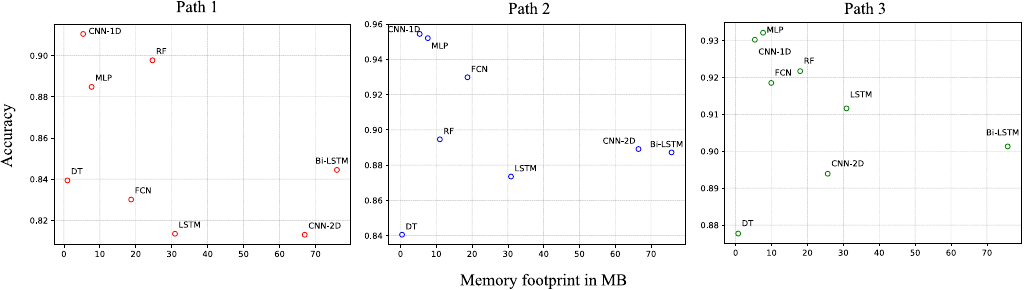}
  \caption{Memory footprint (in MB) against accuracy of the ML models,
    for all three paths.}
  \label{fig:acc_vs_memory}
\end{figure*}

\begin{figure*}[ht]
  \centering
  \includegraphics[scale=0.95]{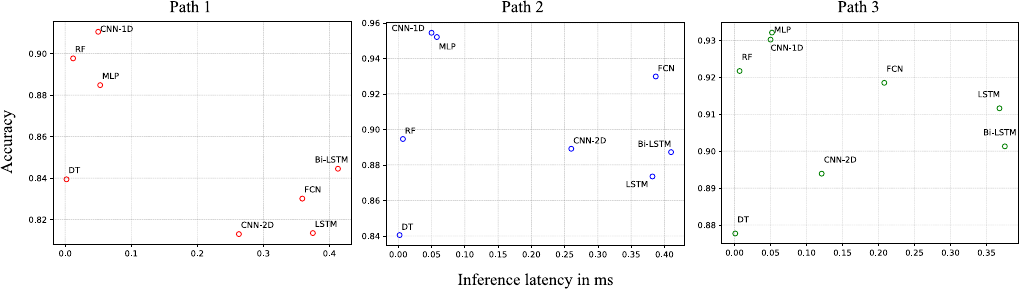}
  \caption{Inference latency (in ms) against accuracy of the ML
    models, for all three paths.}
  \label{fig:acc_vs_infertime}
\end{figure*}

\begin{figure*}[ht]
  \hspace{2mm} \centering
  \includegraphics[scale=0.95]{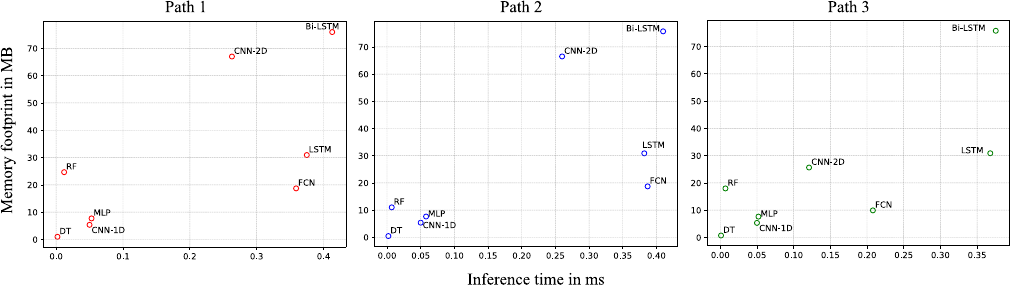}
  \caption{Inference latency (in ms) against memory footprint (in MB)
    of the ML models, for all three paths.}
  \label{fig:memory_vs_infertime}
\end{figure*}

\subsubsection{Inference Latency and Throughput}
In this work inference latency and throughput estimations are
performed on an Intel Core i5-10210U CPU at a base clock speed of 1.60
GHz, which has Intel Turbo Boost Technology for increasing the speed
up to 2.10 GHz under heavy load. The device is equipped with 16 GB
memory and Windows 10 operating system. Hardware with such
specifications typically represents the high end of the currently
available range of edge devices, in terms of performance. The device
is restarted prior to each inference run, and only the inferencing
application is allowed to execute in JupyterLab. The size of the
dataset for inferencing is maintained at 100,000 for all models
tested, i.e., each model is given a batch of 100,000 inputs, with each
input having a length of 30 timesteps of input features, for
prediction. The total time consumed for this amount of predictions is
averaged to get the time per single prediction in milliseconds (ms).
Likewise, the test is performed 10 times for each model, and the
average and standard deviation are calculated.

Tables \ref{table:inference_latency} and
\ref{table:inference_throughput} indicate that DT model is the
fastest, clearing the other models by a great margin, across all three
paths, followed by RF. The next fastest are CNN-1D and MLP models
respectively, with a very narrow margin separating each other. CNN-2D
model, follows thereafter, succeeded by FCN and LSTM models depicting
similar latency performances. BiLSTM model is clearly the slowest
compared to the others.

%0.04167s
Out of the sensors used in the use case scenario, the IMU sensor is
the fastest, sampling data at every 0.041s ($\sim$24Hz). All evaluated
models are capable of meeting this frequency demand. This, however, is
considering the inference latency alone, with the overheads such as
feature processing and other processing latencies neglected. MLP, FCN,
DT, RF, LTSM, BiLSTM, CNN-1D and CNN-2D can cater to sensor sampling
rates up to 14kHz, 2kHz, 5MHz, 172kHz, 2kHz, 2kHz, 18kHz, 3kHz
respectively, at their slowest. Considering the application at hand,
all these inference rates are faster than the fastest sensor used.
Hence, from the inference latency and throughput perspective, using
the hardware on which the experiments are run, all the models are
deployable in the use case scenario described in this paper.
%!!! This is distinctly faster that GPS or vision-based systems.

Apart form the model architectures, the inference latency and
throughput substantially depends on the hardware characteristics
including the processor speed, memory, operating system if any used,
whether any accelerator hardware used and how well the hardware is
optimized for deep learning applications.
% However, it is highly unlikely that the order of the sequence could change, however the difference of values might change depending on the hardware.

\begin{table*}[htbp]
  \caption{Inference latency (milliseconds) of different ML models
    with standard deviation, approximated to 3 decimal places (except
    for DTs). The lowest inference latency for each path is
    highlighted.
%		The tests are performed with a batch size of 100,000 for 10 times for each model, and 
  }
  \begin{center}\scriptsize
    \begin{tabular}{|c|c|c|c|c|c|c|c|c|}
      \hline
      \textbf{Path}&\textbf{MLP}&\textbf{FCN}&\textbf{DT}&\textbf{RF}&\textbf{LSTM}&\textbf{BiLSTM}&\textbf{CNN-1D}&\textbf{CNN-2D}\\
      \hline
       1&0.053$\pm$0.003&0.359$\pm$0.012&\textbf{0.002$\pm$0}&0.012$\pm$0.003&0.375$\pm$0.009&0.413$\pm$0.028&0.05$\pm$0.005&0.263$\pm$0.008\\
      \hline
       2&0.058$\pm$0.009&0.387$\pm$0.012&\textbf{0.002$\pm$0}&0.007$\pm$0.001&0.382$\pm$0.017&0.41$\pm$0.03&0.050$\pm$0.001&0.26$\pm$0.011\\
      \hline
       3&0.052$\pm$0.010&0.208$\pm$0.016&\textbf{0.001$\pm$0}&0.007$\pm$0&0.368$\pm$0.009&0.3755$\pm$0.007&0.05$\pm$0.005&0.121$\pm$0.013\\
      \hline
    \end{tabular}
    \label{table:inference_latency}
  \end{center}
\end{table*}

\begin{table*}[htbp]
  \caption{Inference throughput of the ML models in predictions per
    ms, with the standard deviation, approximated to the 2 decimal
    places. The highest throughput values corresponding to each path
    is highlighted. }
  \begin{center}\scriptsize
    \begin{tabular}{|c|c|c|c|c|c|c|c|c|}
      \hline
      \textbf{Path}&\textbf{MLP}&\textbf{FCN}&\textbf{DT}&\textbf{RF}&\textbf{LSTM}&\textbf{BiLSTM}&\textbf{CNN-1D}&\textbf{CNN-2D}\\
      \hline
       1&19.03$\pm$1.37&2.79$\pm$0.09&\textbf{569.37$\pm$77.39}&91.09$\pm$24.69&2.67$\pm$0.07&2.43$\pm$0.15&20.29$\pm$2.19&3.81$\pm$0.11\\
      \hline
       2&17.46$\pm$2.24&2.58$\pm$0.08&\textbf{644.57$\pm$35.19}&136.45$\pm$9.27&2.62$\pm$0.11&2.45$\pm$0.16&19.92$\pm$0.46&3.85$\pm$0.16\\
      \hline
       3&19.98$\pm$3.24&4.83$\pm$0.37&\textbf{1080.17$\pm$114.97}&147.55$\pm$8.49&2.72$\pm$0.06&2.66$\pm$0.05&20.24$\pm$1.59&8.37$\pm$0.92\\
      \hline
    \end{tabular}
    \label{table:inference_throughput}
  \end{center}
\end{table*}

\subsubsection{Correlation between the Metrics}

In average, four clusters of models can be identified in terms of
memory footprint versus accuracy from figure \ref{fig:acc_vs_memory}.
DTs alone form a cluster having the lowest memory footprint but
significantly lower accuracy than the other models for paths 2 and 3.
CNN-1D and MLP are clustered together, with low memory footprint and
high accuracy (high-accuracy-cluster). For all three paths, LSTM
consistently represents a cluster with mid-range for both memory
footprint and accuracy (mid-range-cluster). For paths 1 and 2, BiLSTM
and CNN-2D models form a cluster for paths 1 and 2, having higher
memory footprints but lower accuracy than MLP and CNN-1D. For path 3,
however, BiLSTM has a significantly higher memory footprint compared
to CNN-2D, forming a cluster by it's own. CNN-2D in this case joins
the mid-range-cluster. RF and FCN mostly represent the high-accuracy
cluster, while alternating to mid-range-cluster for path 2 (RF) and
path 1 (FCN), respectively.

Figure \ref{fig:acc_vs_infertime} illustrates inference latency versus
accuracy. The models MLP and CNN-1D demonstrate high accuracy and low
inference latency consistently in all three paths. The DT and RF
models relatively have the lowest latencies, with the latter
alternating between high to mid-range accuracies, in contrast to DT
which alternate between mid to low range. For paths 1 and 2, the
CNN-2D, FCN, LSTM, and BiLSTM models represent a cluster with
comparatively higher inference latency, with the majority in mid-range
accuracy. The most distinct exceptions is FCN for path 2 which has a
higher accuracy. This cluster evolves slightly for path 3 with the FCN
and CNN-2D models shifting towards the mid-range latency in contrast
to paths 1 and 2.

Memory footprint can be used as an indicator of the scale and
complexity of a model. Figure \ref{fig:memory_vs_infertime}, inference
latency versus memory footprint, clearly verifies this relationship,
which indicates that, as the memory footprint of neural network-based
models increases, the inference latency also increases, indicating an
increase in model complexity.
%RF models lie in the mid-range of memory footprint while possessing low inference latencies. 
DT, MLP and CNN-1D models are concentrated at the low memory footprint
and low inference latency range. RF models positions itself close to
this cluster, but maintaining a notable space from the rest in paths 1
and 3. Models such as FCN, CNN-2D, LSTM, and BiLSTM spread from the
mid to high range in terms of both inference latency and memory
footprint.

\subsection{Discussion and Future Improvements}
One of the potential challenges commonly associated with ML-based
methods is the occurrence of data drift. This phenomenon can emerge
either gradually over time in sensors or due to variations in the
positioning environment, such as changes in the factory assembly line
setup. In such cases, the models need to be retrained using new data.
This highlights a notable drawback of the proposed approaches when
compared to the majority of existing IPSs, whose position estimation
does not solely rely on ML models. The same limitation holds true in
cases when the proposed method requires scaling up.

The generality of the indoor positioning problem addressed in this
work, is restricted to the asset's moving along a predetermined path
at randomly varying speeds, with varying times of path completion, and
with forward and return journeys lying along the same path. To improve
this further, we plan to enhance the models' capabilities to localize
accurately even in the presence of anomalies within the environment.
This will involve capturing data instances that encompass various
scenarios, including instances where motion deviates from the
predefined track exposing the models to unseen data, moments of pause
at random locations for varying durations, collisions with nearby
structures and other unexpected events. Thereby, we intend to enhance
the generality of the presented algorithms.

Furthermore, we intend to expand the generalizability of the evaluated
ML models to characterize and classify common motions occurred in
indoor environments such as moving along ramps, elevators, conveyor
belts, left and right turns, etc. This broadens their applicability on
new paths with transfer learning, requiring smaller training datasets,
accelerated training and improved accuracy. From the results, we see
MLP and CNN-1D which performs overall the best, consistently for all
three paths, are already suitable for this application.

Features such as temperature, humidity, and spectral attributes
exhibit seasonality effects that can potentially influence the
outcomes of the classification. Nevertheless, our dataset lacks the
representation of these seasonal effects. Consequently, the current
version of this study does not address accuracy variations across
different seasons and will be investigated in future works.

We further plan to investigate the performance of the models by
deploying them on low power and low performance edge devices
preferably on microcontrollers, while sensing data in real time,
on-site.

%\subsection{Future improvements}
%The generality of the indoor positioning problem addressed in this work, is restricted to the asset's moving along a predetermined path at randomly varying speeds, with varying times of path completion, and with forward and return journeys lying along the same path. To improve this, our future work will include paths with anomalies such as having longer, random pauses along the traversal, going out of the predetermined path, and detecting the correct branch when there are several branches available in a given main path. Thereby, we intend to enhance the generality of the presented algorithms.

\section{Conclusion}
\label{sec:conclusion}
In this paper, we have presented for the first time, an indoor
positioning method using ML, by fusing motion and ambient sensors,
operating independently from any external infrastructure. We also
introduce the novel Motion-Ambient dataset which includes multivariate
time series data. Using this dataset, we model the indoor positioning
problem as an MTSC problem and formulate, train and evaluate ML models
based on the architectures DT, RF, LSTM, BiLSTM, CNN-1D, and CNN-2D.
These are compared and analysed against the standard MTSC benchmarking
algorithms, MLP and FCN \cite{Wang_2017} using the comparison metrics,
accuracy, loc-score, memory footprint, inference latency and
throughput. The results show that all the models achieve accuracies
higher than 80\%, which is in general sufficiently precise for the
type of use cases that we address in this work. All the models meet
the latency demands considering the application in hand. The memory
footprint however ranges approximately from 0.5 - 76 MB. CNN-1D
demonstrates the most balanced performance, closely followed by MLP.
It is also worth noting that DT and RF, having the best performance in
memory footprint and inference latency, can be made much superior if
optimized to enhance accuracy through feature engineering manually.
Furthermore, we can conclude that ML can be used to solve the indoor
positioning problem. We intend to deploy our system in a real factory
environment.

\bibliographystyle{plainurl}
\bibliography{bibliography}
\end{document}